\documentclass[aps,prl,showpacs,amsmath,preprint,superscriptaddress,floatfix,nobalancelastpage,footinbib]{revtex4-2}

\usepackage[colorlinks=true, citecolor=black, linkcolor=black, urlcolor=black]{hyperref}
\usepackage[all]{hypcap}
\usepackage{amsmath}
\usepackage{enumerate,bm}
\usepackage{graphicx}
\usepackage{color}
\usepackage{esint}
\usepackage[usenames,dvipsnames]{xcolor}
\usepackage{subfigure}
\usepackage{flushend}
\usepackage{amssymb}
\usepackage{float}
\usepackage{dcolumn}
\usepackage{ulem} 

\usepackage{setspace}
\usepackage{verbatim}

\begin{document}

\author{Dongbin~Shin}
\email{dongbin.shin@mpsd.mpg.de}
\affiliation{Max Planck Institute for the Structure and Dynamics of Matter and Center for Free Electron Laser Science, 22761 Hamburg, Germany}

\author{Simone~Latini}
\affiliation{Max Planck Institute for the Structure and Dynamics of Matter and Center for Free Electron Laser Science, 22761 Hamburg, Germany}

\author{Christian~Sch\"afer}
\affiliation{Max Planck Institute for the Structure and Dynamics of Matter and Center for Free Electron Laser Science, 22761 Hamburg, Germany}

\author{Shunsuke~A.~Sato}
\affiliation 
{Center for Computational Sciences, University of Tsukuba, Tsukuba 305-8577, Japan}
\affiliation{Max Planck Institute for the Structure and Dynamics of Matter and Center for Free Electron Laser Science, 22761 Hamburg, Germany}

\author{Umberto~De~Giovannini}
\affiliation{Max Planck Institute for the Structure and Dynamics of Matter and Center for Free Electron Laser Science, 22761 Hamburg, Germany}
\affiliation{Nano-Bio Spectroscopy Group,  Departamento de Fisica de Materiales, Universidad del País Vasco UPV/EHU- 20018 San Sebastián, Spain}

\author{Hannes~H\"ubener}
\affiliation{Max Planck Institute for the Structure and Dynamics of Matter and Center for Free Electron Laser Science, 22761 Hamburg, Germany}

\author{Angel~Rubio}
\email{angel.rubio@mpsd.mpg.de}
\affiliation{Max Planck Institute for the Structure and Dynamics of Matter and Center for Free Electron Laser Science, 22761 Hamburg, Germany}
\affiliation{Nano-Bio Spectroscopy Group,  Departamento de Fisica de Materiales, Universidad del País Vasco UPV/EHU- 20018 San Sebastián, Spain}
\affiliation{Center for Computational Quantum Physics (CCQ), The Flatiron Institute, 162 Fifth avenue, New York NY 10010.}

\title{The quantum paraelectric phase of SrTiO$_3$ from first principles}


\date{\today}
\begin{abstract}
We demonstrate how the quantum paraelectric ground state of SrTiO$_3$ can be accessed via a microscopic \textit{ab initio} approach based on density functional theory. 
At low temperature the quantum fluctuations are strong enough to stabilize the paraelectric phase even though a classical description would predict a ferroelectric phase. 
We find that accounting for quantum fluctuations of the lattice and for the strong coupling between the ferroelectric soft mode and lattice elongation is necessary to achieve quantitative agreement with experimental frequency of the ferroelectric soft mode.
The temperature dependent properties in SrTiO$_3$ are also well captured by the present microscopic framework.
\end{abstract}

\maketitle

SrTiO$_3$ is arguably one of the most intensively studied materials of the perovskite family~\cite{Schooley1964,Koonce:1967fe,Muller:1979je,Haeni2004,Itoh:1999eq,Li2019,Nova2019}. 
Under ambient conditions SrTiO$_3$ is an insulating paraelectric, however a transition to ferroelectric phase can be induced by numerous mechanism with relatively low activation energy. 
The temperature-strain phase diagram of SrTiO$_3$ is, at low temperatures, characterized by a very small region of paraelectricity with a ferroelectric phase emerging for small strains~\cite{Li:2006}. 
With increasing temperature the paraelectric region widens, yet, ferroelectricity remains accessible even above room temperature for high enough strain~\cite{Haeni2004}. 
The proximity of the ambient paraelectric phase to a ferroelectric phase in the phase diagram is also underlined by the possibility to induce the phase transition through oxygen isotope substitution~\cite{Itoh:1999eq} or by applying an intense laser pulse~\cite{Li2019,Nova2019}.  

At low temperature ($T<0.5$~K), SrTiO$_3$ displays a striking superconductive behavior with low carrier concentration~\cite{Schooley1964,Appel1966}.
Notably, this superconducting phase is characterized by the competition of isotope oxygen doping and quantum fluctuation of the transverse optical (TO) soft phonon mode. 
Both compete for the formation of electron pairing and ferroelectricity, a mechanism known as quantum criticality~\cite{Rowley2014,Edge:2015fj,Rischau2017}. 
The microscopic details of the superconducting phase and quantum criticality in SrTiO$_3$ are still debated~\cite{Narayan2019,Itahashi2020}.
These quantum fluctuations, however, play an important role even for the ground state, where they are decisive in a competition between ferro- and paraelectricity.

Its low temperature paraelectric phase makes SrTiO$_3$ stand out among the other compounds of the ABO$_3$ perovskites family such as BaTiO$_3$ and PbTiO$_3$~\cite{Shirane1952,Samara1971,Zhong1995,Bellaiche2000,Guo2000,Miyasaka2003,Haeni2004,Fennie2006,Lee2010} where, with decreasing temperature, the polar TO phonon softens until it turns unstable and the material becomes ferroelectric. 
This phase is characterised by a double well shape of the potential energy landscape along the coordinate of the so-called ferroelectric soft (FES) mode, such that at zero temperature the ground state is in a degenerate superposition of positively and negatively polarised states, which endows the material with an internal macroscopic polarization when the degeneracy is lifted by spontaneous symmetry breaking.
While the FES mode of SrTiO$_3$ displays a similar characteristic softening, it stabilizes at low temperatures and no ferroelectricity is observed. 
This low temperature behaviour has been rationalized in terms of quantum fluctuations that prevent the formation of a macroscopic dipole~\cite{Muller:1979je}.
This phenomenon of quantum paraelectricity and has been invoked to explain the behavior of other complex oxides~\cite{Muller:1979je,Rytz1980,Akbarzadeh2004,Zhang2020}.
A number of models have been developed to describe the dielectric properties of this phase, in particular the Barrett and Vendik models have been widely used to understand quantum paraelectric behavior~\cite{Cowley1964,Muller:1979je,Rytz1980,Westwanski1999,Marques2005,Fujishita2016}. 
The decisive role played by quantum fluctuations for the temperature dependent competition between the ferro- and paraelectric phases has been confirmed by quantum Monte Carlo calculations with an effective Hamiltonian for phenomenologically strained SrTiO$_3$~\cite{zhong1996,Martonak1996} as well as for other materials~\cite{Akbarzadeh2004,Zhang2020}.
The quantum paraelectric phase is therefore now widely accepted as the explanation for the low temperature behaviour of SrTiO$_3$~\cite{Muller:1979je,Song1996,He2020}, however the ground state of the quantum paraelectric phase and the frequency of the FES mode at low temperature have not yet been described by a microscopic theory.

The conventional first principles method to evaluate phonon frequencies of materials at zero temperature is density functional perturbation theory (DFPT)~\cite{Gonze1997}.
In DFPT calculations, which are based on a harmonic description of the lattice, an imaginary eigenvalue of the dynamical matrix indicates a phonon instability that can point towards a phase transition. 
However, DFPT does not include the quantum nuclear effects that are believed to stabilise the paraelectric phase in SrTiO$_3$ and therefore wrongly predicts an instability of the FES mode that suggests a phase transition to ferroelectrity~\cite{Aschauer:2014,Zhou:2018,Wahl:2008}.
This failure to describe quantum paraelectricity and its influence on the frequency of the FES mode, prevents the application of standard \textit{ab initio} methods to give a microscopic explanation of recent experiments on quantum criticality and THz induced ferroelectricity in SrTiO$_3$~\cite{Rowley2014,Edge:2015fj,Rischau2017,Li2019,Nova2019}.

\begin{figure}[t!]
  \centering
  {\includegraphics[width=0.7\textwidth]{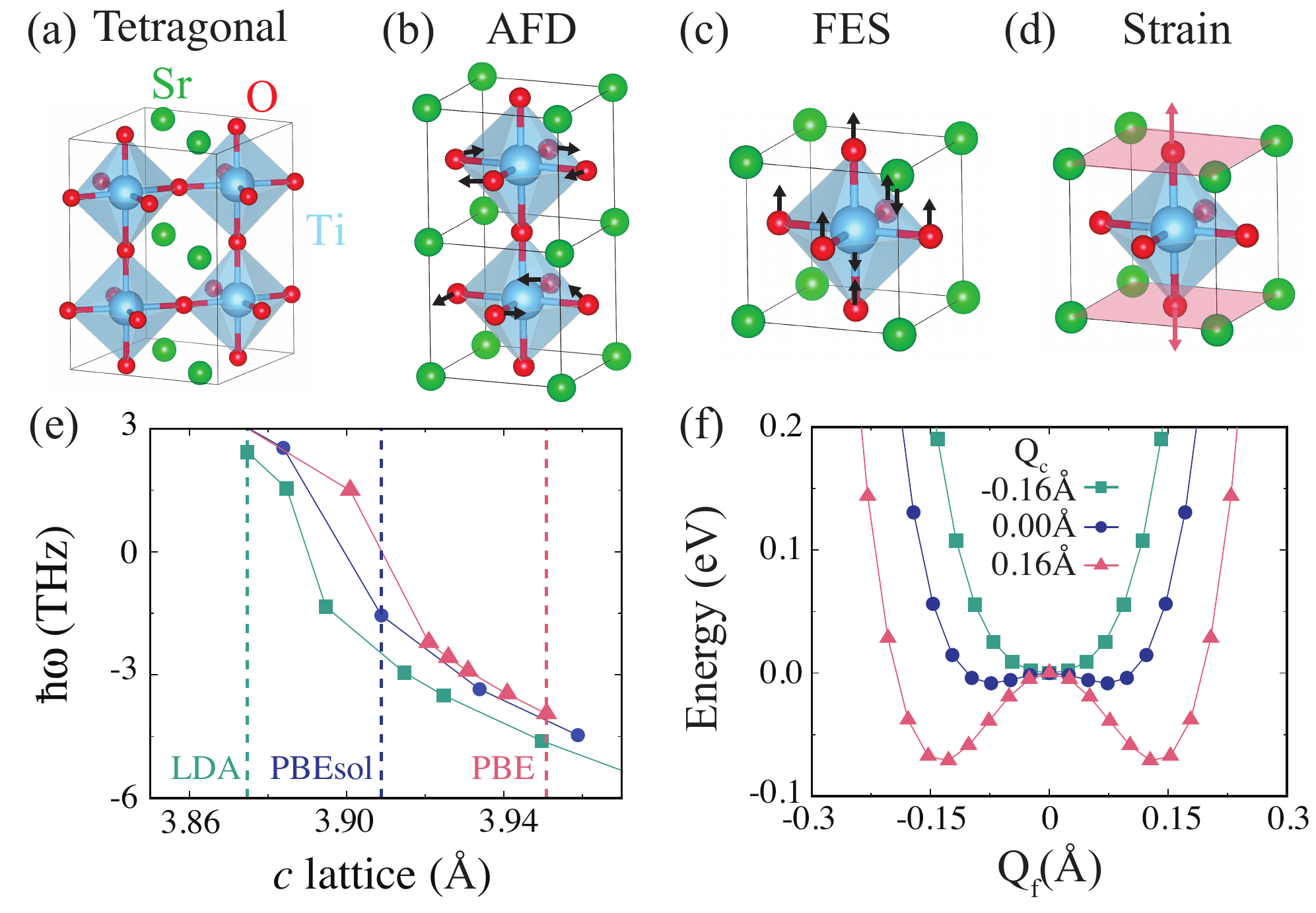}}
   \caption{(a) Atomic geometry of SrTiO$_3$ in the tetragonal phase. Eigenvector of (b) antiferro-distortive mode and (c) ferroelectric soft mode. (d) Strain direction along $c$-axis. (e) Frequency of the FES mode as a function of the lattice parameter $c$ predicted by DFTP within LDA, PBE, and PBEsol functionals. (f) Cuts of the PBE potential energy surface of the FES mode for different lattice variation $Q_{\rm{c}}$. In (e), vertical dashed lines indicate the optimized lattice constants for each functional.}
\end{figure}

In this letter, we unambiguously confirm the quantum nuclear nature of the ground state of SrTiO$_3$ based on DFT calculations and show that quantum fluctuations of the FES mode stabilise the paraelectric phase at low temperatures. To describe the quantum behavior of the lattice dynamics, we compute the potential energy surface obtained from DFT and construct a lattice-nuclear Schr\"odinger equation. 
We find that it is, indeed, not enough to only describe the FES mode as a quantum state in a 1D Schr\"odinger equation, but the nonlinear coupling to the lattice needs to be included in the quantum description. 
From this description we correctly reproduce frequency of the FES mode at zero temperature as well as the temperature dependence of the frequency and dielectric constant in the quantum pararelectric phase, which well agree with the experimental observations. 
We furthermore show that the crystal properties obtained with DFT strongly depend on the exchange-correlation functional and that a correct description of both the lattice constants as well as the atomic positions are crucial to obtain the correct quantum paraelectric phonon energy. 

\begin{table}[t!]
\caption{\label{tab:table1}%
The lattice parameter of tetragonal SrTiO$_3$ and the associated FES mode frequency for different functionals. The functional-dependent lattice parameters are consistent with previously reported results~\cite{Wahl:2008}.}
\begin{ruledtabular}
\begin{tabular}{cccccc}
\multicolumn{1}{c}{\textrm{}}&
\multicolumn{1}{c}{\textrm{LDA}}&
\multicolumn{1}{c}{\textrm{PBEsol}}&
\multicolumn{1}{c}{\textrm{PBE}}&
\multicolumn{1}{c}{\textrm{HSE06}}&
\multicolumn{1}{c}{\textrm{Exp.}}\\
\hline
a (\AA) & 3.843 & 3.882 & 3.929 & 3.908 & 3.898~\cite{Cao2000} \\
a/c & 1.008 & 1.007 & 1.005 & 1.004 & 1.001~\cite{Heidemann1973}\\
AFD ($^\circ$) & 6.4 & 6.1 & 5.3 & 3.2 & 2.1~\cite{Unoki1967}\\
DFPT (THz) & 2.4 & -1.5 & -3.9 & -2.2~\cite{Wahl:2008}  \\
\end{tabular}
\end{ruledtabular}
\end{table}

Below $105$~K, the crystal structure of SrTiO$_3$ forms a tetragonal unit cell with the oxygen octahedra rotated with respect to the cubic cell.
This rotation counteracts the formation of ferroelectricity and is hence usually referred to as an \textit{anti-ferro}-distortive (AFD) motion~\cite{Aschauer:2014}. 
Therefore, the tetragonal geometry can be described as a $\sqrt{2} \times \sqrt{2} \times 2$ supercell of the primitive cubic perovskite ABO$_3$ unit cell with an additional AFD rotation, as depicted in Figs.~1(a) and 1(b). 
This AFD in-plane rotation is accompanied by an elongation and a contraction of the $c$ and $a$ lattice vectors relative to the cubic structure~\cite{Aschauer:2014}.
To investigate the optimized geometry and total energy, we perform DFT calculations using the Quantum Espresso package~\cite{Giannozzi2017}.
The projector augmented wave method is employed to describe core level atomic orbitals and a plane-wave basis set with 70 Ry energy cut-off is used.
The Brillouin zone is sampled with $6 \times 6 \times 4$ $\mathbf{k}$-points.
In Tab.~1, we summarize the lattice parameter $a$, $c/a$ ratio and AFD rotation angle obtained with various DFT functionals. 
Comparing with experimental observation~\cite{Loetzsch2010,Cao2000,Heidemann1973,Unoki1967}, local density approximation (LDA)~\cite{Perdew1992} and Perdew-Berke-Ernzerhof revised for solid (PBEsol)~\cite{Csonka2009} functionals provide a contracted lattice parameter $a$ and a higher $a/c$ ratio with an over-rotated AFD angle. 
Even though the Perdew-Berke-Ernzerhof (PBE)~\cite{Perdew1996} functional and Heyd-Scuseria-Ernzerhof (HSE06)~\cite{Heyd2003,Krukau2006} hybrid functional describe elongated $a$ and $c$ lattices with over-rotated AFD angle, these lattice parameters are closer to experimental observations than the former two functionals.

By performing DFPT calculations on SrTiO$_3$ we evaluated the FES mode energy for different functionals and found that for the respective optimized lattice parameters (indicated by the vertical dashed lines), the FES mode is unstable for PBE and PBEsol but not for LDA, see Fig.~1(e) and Tab.~1. 
This indicates that LDA would predict a classical paraelectric ground state in contrast with PBE and PBEsol (as well as hybrid functionals~\cite{Wahl:2008,El-Mellouhi2011}) predicting a phase transition. 
Regardless of the functional an additional frequency softening of the FES mode is induced by the $c$ lattice parameter elongation. 
This instability can be understood by considering the potential energy landscape of the FES mode computed with PBE.
Given the strong dependence of the DFPT results on the lattice parameter $c$ we calculated the potential energy surface for different $c$ values, denoted by $Q_{\rm{c}}$, as reported in Fig.~1(f).
The potential energy surface for the FES mode is evaluated by displacing the atomic positions along the optimized ferroelectric geometry with respect to the optimized tetragonal geometry; we denote this parameterized displacement as $Q_{\rm{f}}$. 
At the optimized lattice parameter, $Q_{\rm{c}}=0$~\AA , the potential energy surface shows a shallow double well potential. 
This explains the instability found in DFPT and why in absence of quantum fluctuations the system would spontaneously collapse into one of the two wells inducing a ferroelectric polarisation.
Similar potential energy calculations for the LDA functional (not shown) provide a single well dispersion and hence no phase transition is expected. 
Changing $Q_{\rm{c}}$ results in an asymmetric behaviour when using the PBE functional: for negative $Q_{\rm{c}}$, i.e. lattice contraction, the double well disappears while it deepens under lattice expansion, positive $Q_{\rm{c}}$. 
This behavior explains why the phase transition between the ferroelectric and paraelectric phases of SrTiO$_3$ can be easily induced by strain~\cite{Haeni2004,Antons2005}.

\begin{figure}[t!]
  \centering
  {\includegraphics[width=0.7\textwidth]{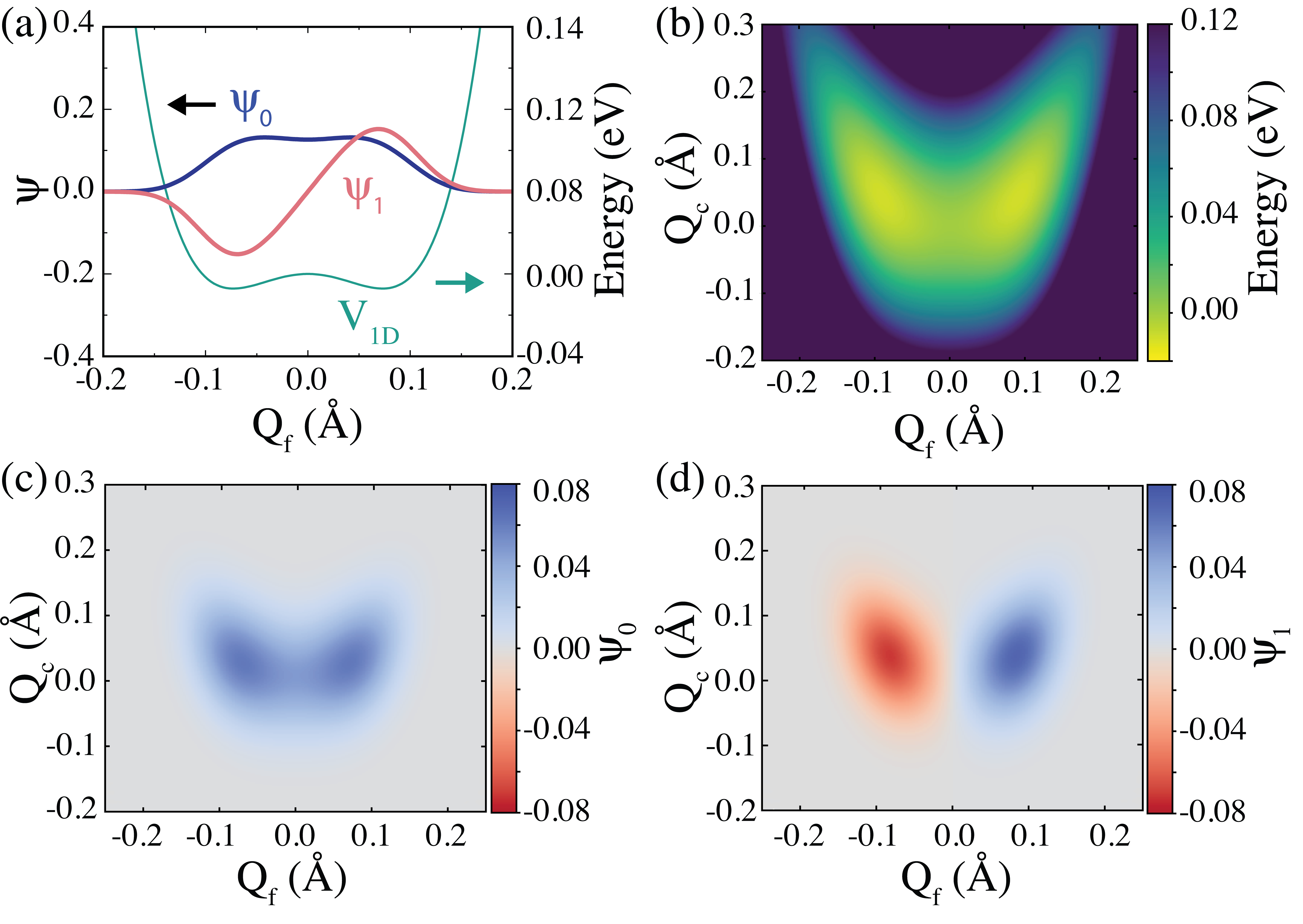}}
   \caption{(a) Potential energy surface ($V_{\text{1D}}$) and ground and 1st excited wavefucntions obtained by solving 1DSE. (b) 2D Potential energy surface ($V_{\text{2D}}^{\text{FES,c}}$) consists of FES mode and $c$ lattice. (c) Ground and (d) 1st excited wavefunctions obtained by solving 2DSE.}
\end{figure}

\begin{table}[t!]
\caption{\label{tab:table2}%
Computed frequency of the FES mode by solving a nuclear Schr\"odinger equation with Wentzcovitch-type fictitious cell mass and experimentally observed frequency of FES mode in THz units.}
\begin{ruledtabular}
\begin{tabular}{cccccc}
\multicolumn{1}{c}{\textrm{in THz}}&
\multicolumn{1}{c}{\textrm{LDA}}&
\multicolumn{1}{c}{\textrm{PBEsol}}&
\multicolumn{1}{c}{\textrm{PBE}}&
\multicolumn{1}{c}{\textrm{HSE06}}&
\multicolumn{1}{c}{\textrm{Exp.}}\\
\hline
1DSE& 4.4 & 2.8 & 1.2 & 1.5 &    \\
2DSE& 4.1 & 2.4 & 0.44 & 0.83 &   \\
Neutron &  & & & & 0.53 (5K)~\cite{Shirane1969}  \\
Hyper-Raman & & & & & 0.49 (6K)~\cite{Yamanaka2000} \\
 & & & & & 0.53 (9K)~\cite{Vogt1995} \\
\end{tabular}
\end{ruledtabular}
\end{table}
 
Given the shallow double well found for the FES mode in SrTiO$_3$, it is necessary to include quantum-nuclear effects, namely the zero-point motions of the atoms. 
We restrict our description of the lattice dynamics to the FES mode and the lattice motion in the $c$-direction, which we have established above to be intimately dependent.
We sample the DFT total energy for $25 \times 13$ geometries along the FES mode parameterized by $Q_{\rm{f}}$ and the lattice expansion parameterized by $Q_{\rm{c}}$. 
We then fit the potential energy surface; details on the definition of the FES eigenvector and the fitting coefficients are reported in Supplementary Material~\cite{SM}.
 
First we solve the nuclear Schr\"odinger equation in 1D (1DSE) for the FES mode along the potential energy curve for the optimized lattice parameters ($Q_{\rm{c}}=0$). 
The Hamiltonian for the 1DSE reads: $\hat{H}_{\text{1D}}^{\text{FES}}=\frac{\hat{P}^2_f}{2M_f}+\sum_{i=1}^6 k_{f,i}\hat{Q}^{2i}_f$, where $\hat{P}_f$ and $M_f=1.76 \times 10^{-25}$~kg are the momentum operator and the FES phonon mass, respectively and $k_{f,i}$ are the coefficients that parameterize the DFT potential energy surface.
While for a ferroelectric one would expect a double degenerate ground state in the double well potential, the diagonalization of the 1DSE provides a non-degenerated ground ($\psi_0$) and 1st excited ($\psi_1$) state, which are depicted in Fig.~2(a). 
The energy difference between the 1st excited and the ground state ($\hbar \omega = \epsilon_1 - \epsilon_0$) can be identified as the FES phonon frequency; the values are summarized in Tab.~2 for different functionals.
Except for LDA, where the quantum effects in the 1DSE entails only a frequency stiffening, the FES frequency changes sign due to quantum fluctuations.
While a positive frequency for the FES mode correctly indicate that the paraelectric phase is stable, the values are too high compared to the experiments.

To obtain the correct low temperature FES mode frequency, we find it is necessary to explicitly include its coupling to the lattice mode $Q_{\rm{c}}$ in a two-dimensional (2D) lattice-nuclear Schr\"odinger equation (2DSE). 
The corresponding Hamiltonian given as $\hat{H}_{\text{2D}}^{\text{FES,c}}={\hat{P}^2_f}/{2M_f}+{\hat{P}^2_c}/{2M_c}+\hat{V}_{2D}^{\text{FES,c}}$ is built on the 2D potential energy surface, as shown in Fig.~2(b) and calculated as described above in terms of $\hat{Q}_f$ and $\hat{Q}_c$: $\hat{V}_{\text{2D}}^{\text{FES,c}}=\sum_{i=1}^6 k_{f,i}\hat{Q}^{2i}_f + \sum_{j=2}^5 k_{c,j}\hat{Q}^{j}_c+\sum_{i=1}^6\sum_{j=1}^5 k_{fc,i,j}\hat{Q}^{2i}_f \hat{Q}^{j}_c $.
The total cell mass ($M_{tot}=\sum_i M_i=1.22\times 10^{-24}$~kg) and Wentzcovitch-type fictitious cell mass ($\frac{3M_{tot}}{4\pi^2\Omega^{2/3}}=9.34 \times 10^{-26}$~kg)~\cite{Wentzcovitch1991} are both considered for the mass of the lattice ($M_c$); we verified that the deviation of the FES frequency due to the choice of the lattice mass is less than $5\%$.
The ground ($\psi_0$) and 1st excited states ($\psi_1$), obtained from the solution of the 2DSE are depicted in Figs.~2(c) and 2(d).
The characteristic node of the 1st excited state along the $Q_{\rm{f}}$ indicates that the state is of FES mode character and hence can be used to determine the FES frequency.
Similar to the 1DSE case, the 2DSE provides non-degenerate ground and 1st excited states with positive FES phonon frequencies for all the investigated functionals, as summarized in Tab.~2.
All FES mode frequencies become softer, when the FES-lattice interaction is included, as compared to the values obtained by only using the 1DSE.
Importantly, the FES phonon frequencies evaluated by PBE ($0.44$~THz) and HSE06 ($0.83$~THz) are close to the experimentally measured values at low temperature~\cite{Shirane1969,Vogt1995,Yamanaka2000}.
We conclude that both quantum fluctuations and the FES-lattice interactions are crucial in determining the frequency of the FES mode.

We then extend our microscopic approach to include the effect of finite temperatures.
Experimentally it has been shown that a flat temperature dependence of the FES mode is expected in the quantum paraelectric phase ($T<4$K), \cite{Muller:1979je,Song1996} whereas for increasing temperatures a stiffening of the FES mode and a drop of the dielectric function are observed~\cite{Vogt1995,Song1996,Yamanaka2000,Li2019}.
We first evaluate the temperature dependent FES frequency via \textit{ab initio} molecular dynamics simulations with a thermostat~\cite{Wentzcovitch1991}.
As shown in Fig.~3(a), the results with PBE provide values which are comparable with the experimental observations at high temperature, where the quantum effects are overshadowed by thermal fluctuations (see Supplementary Materials for details~\cite{SM}).
This indicates that PBE provides a realistic potential energy surface and that the effect of a slightly overestimated lattice is negligible~\cite{Zhou:2018}.
To include temperature in our quantum lattice model we apply Kubo's formula for the linear response of a thermal state to a perturbation $\hat{H}'(t) = -Z^{*}\hat{Q_{\rm{f}}} E(t)$, where $Z^{*}$ is the FES mode effective charge that we assume to be temperature independent. 
The resulting polarizability takes the form: $\alpha(\omega,T)=-\sum_{i,j} \rho_i(T)Z^{*}|D_{ij}|^2 \times (\frac{1}{(\epsilon_j - \epsilon_i)-\omega -i\delta }+\frac{1}{(\epsilon_j - \epsilon_i)+\omega +i\delta }) $, where the dipole matrix and the thermal density matrix are defined as $D_{ij}=\langle \psi_i|\hat{Q}_f|\psi_j\rangle$ and $\rho_i(T)=e^{-(\epsilon_i-\epsilon_0)/k_BT}/\sum_j e^{-(\epsilon_j-\epsilon_0)/k_BT}$, respectively.
The temperature dependent frequency of the FES mode, is then evaluated by averaging over the polarizability as $\omega(T)=\frac{\int \omega Im[\alpha(\omega,T)] d\omega}{\int Im[\alpha(\omega,T)] d\omega}$; the results of this procedure with PBE are depicted in Fig.~3(a).
Even though the 2D potential only includes two degrees of freedom ($Q_{\rm{f}}$ and $Q_{\rm{c}}$), the temperature dependence behavior is well reproduced and the typical flattening at low temperatures ($<10$ K) is evident. 
We assigned the observed deviations of our model from the experiment to the effect of the phonon degrees of freedom that are not included in the model~\cite{Vogt1995,Yamanaka2000,Li2019}. 
An attempt on improving the description of our method by including the effect of temperature on the lattice parameter bearing no significant change to the picture is discussed in Supplementary Material~\cite{SM}.

\begin{figure}[t!]
  \centering
  {\includegraphics[width=0.7\textwidth]{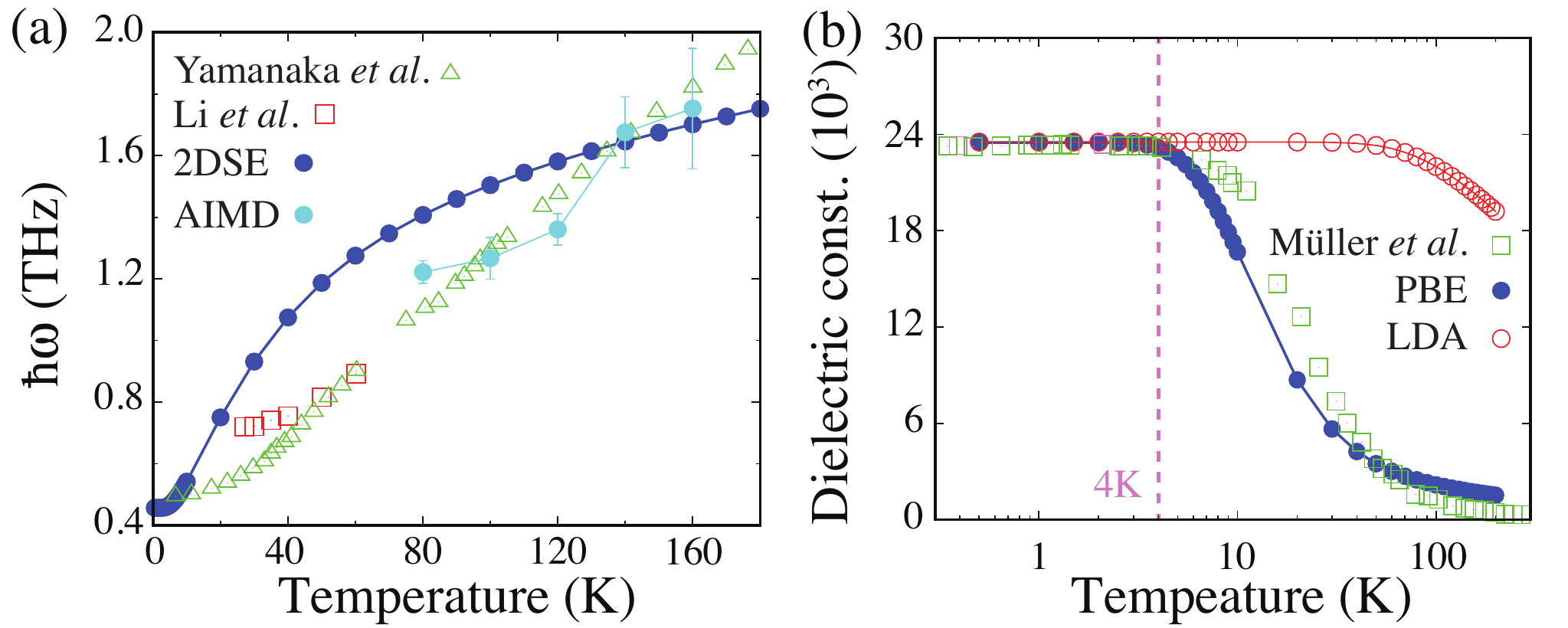}}
   \caption{(a) Temperature dependence of FES mode frequency obtained with PBE and experimental observations~\cite{Yamanaka2000,Li2019}. 
   (b) Temperature dependency of the estimated dielectric constant using Lyddane–Sachs–Teller relation with PBE and LDA functionals and experimental observation~\cite{Muller:1979je}.
   In (b), the vertical dashed line is a guide for eye placed at $4$~K to indicate the dielectric plateau limit.}
\end{figure}

Now, using the Lyddane-Sachs-Teller relation~\cite{Lyddane1941} we can further estimate the temperature dependence of the dielectric function as $\epsilon(T) \sim 1/\omega(T)^2$~\cite{Worlock1967,Muller:1979je,Song1996}.
In Fig.~3(b), we compare with the experimental observed dielectric constants by fixing the value of the dielectric function at zero temperature to the experimental one and using the temperature dependent FES frequency calculated above.
The characteristic flat plateau up to $4$K and the subsequent drop in the dielectric function is reproduced very well by our PBE calculations. 
Similar calculation with LDA do not follow the correct trend, highlighting the failure of LDA at describing the correct potential energy landscape for the FES mode.

In conclusion, we investigated the low temperature quantum behavior of SrTiO$_3$ from a fully microscopic point of view.
In line with the concept of quantum paraelectricity~\cite{Muller:1979je}, we show that only the quantum description of the lattice predicts a stable phonon mode and hence the paraelectric phase; a result in contrast with conventional perturbation theory, which wrongly predicts an instability that leading to a ferroelectric transition at low temperature.  
We show that, not only the quantum fluctuations of the FES phonon stabilize this phase, but that the interaction between FES mode and fluctuations of the $c$ lattice parameter contribute to the ground state and to the first FES eigenmode energy; therefore it is crucial to treat these degrees of freedom on the same footing in order to explain the dynamical properties of SrTiO$_3$. 
Our DFT-based treatment of the lattice-nuclear Schr\"odinger equation provides low temperature phonon frequencies in agreement with the experimentally observed values. 
Combined with thermal statistics, this treatment also reproduces the temperature dependence of the frequency of the FES mode and the dielectric constant, namely the frequency stiffening of the FES mode with increasing temperature and the flat behavior of the dielectric constant stemming from the quantum fluctuation of the lattice. 
Besides providing a detailed and direct first principles description of the FES frequency and the quantum paraelectric ground state of SrTiO$_3$, which has been conjectured for a long time and thus far only been considered in the context of phase dynamics, this work opens new avenues to investigate recently observed light induced ferroelectricity in SrTiO$_3$~\cite{Nova2019,Li2019} and to address the ground state of SrTiO$_3$ embedded in an optical cavity~\cite{Ashida2020,Hannes2020} by providing a model for the low temperature nuclear lattice Hamiltonian.  

\begin{acknowledgments}
We further acknowledge financial support from the European Research Council (ERC-2015-AdG-694097), Grupos Consolidados (IT1249-19), JSPS KAKENHI Grant Number 20K14382, the Cluster of Excellence 'CUI: Advanced Imaging of Matter' of the Deutsche Forschungsgemeinschaft (DFG) - EXC 2056 - project ID 390715994. 
We acknowledge support from the Max Planck–New York Center for Non-Equilibrium Quantum Phenomena.
The Flatiron Institute is a division of the Simons Foundation.
D.S. and S.L. are supported by Alexander von Humboldt Foundation.  
\end{acknowledgments}


\begin{thebibliography}{58}%
\makeatletter
\providecommand \@ifxundefined [1]{%
 \@ifx{#1\undefined}
}%
\providecommand \@ifnum [1]{%
 \ifnum #1\expandafter \@firstoftwo
 \else \expandafter \@secondoftwo
 \fi
}%
\providecommand \@ifx [1]{%
 \ifx #1\expandafter \@firstoftwo
 \else \expandafter \@secondoftwo
 \fi
}%
\providecommand \natexlab [1]{#1}%
\providecommand \enquote  [1]{``#1''}%
\providecommand \bibnamefont  [1]{#1}%
\providecommand \bibfnamefont [1]{#1}%
\providecommand \citenamefont [1]{#1}%
\providecommand \href@noop [0]{\@secondoftwo}%
\providecommand \href [0]{\begingroup \@sanitize@url \@href}%
\providecommand \@href[1]{\@@startlink{#1}\@@href}%
\providecommand \@@href[1]{\endgroup#1\@@endlink}%
\providecommand \@sanitize@url [0]{\catcode `\\12\catcode `\$12\catcode
  `\&12\catcode `\#12\catcode `\^12\catcode `\_12\catcode `\%12\relax}%
\providecommand \@@startlink[1]{}%
\providecommand \@@endlink[0]{}%
\providecommand \url  [0]{\begingroup\@sanitize@url \@url }%
\providecommand \@url [1]{\endgroup\@href {#1}{\urlprefix }}%
\providecommand \urlprefix  [0]{URL }%
\providecommand \Eprint [0]{\href }%
\providecommand \doibase [0]{https://doi.org/}%
\providecommand \selectlanguage [0]{\@gobble}%
\providecommand \bibinfo  [0]{\@secondoftwo}%
\providecommand \bibfield  [0]{\@secondoftwo}%
\providecommand \translation [1]{[#1]}%
\providecommand \BibitemOpen [0]{}%
\providecommand \bibitemStop [0]{}%
\providecommand \bibitemNoStop [0]{.\EOS\space}%
\providecommand \EOS [0]{\spacefactor3000\relax}%
\providecommand \BibitemShut  [1]{\csname bibitem#1\endcsname}%
\let\auto@bib@innerbib\@empty
\bibitem [{\citenamefont {Schooley}\ \emph {et~al.}(1964)\citenamefont
  {Schooley}, \citenamefont {Hosler},\ and\ \citenamefont
  {Cohen}}]{Schooley1964}%
  \BibitemOpen
  \bibfield  {author} {\bibinfo {author} {\bibfnamefont {J.~F.}\ \bibnamefont
  {Schooley}}, \bibinfo {author} {\bibfnamefont {W.~R.}\ \bibnamefont
  {Hosler}},\ and\ \bibinfo {author} {\bibfnamefont {M.~L.}\ \bibnamefont
  {Cohen}},\ }\href@noop {} {\bibfield  {journal} {\bibinfo  {journal} {Phys.
  Rev. Lett.}\ }\textbf {\bibinfo {volume} {12}},\ \bibinfo {pages} {474}
  (\bibinfo {year} {1964})}\BibitemShut {NoStop}%
\bibitem [{\citenamefont {Koonce}\ \emph {et~al.}(1967)\citenamefont {Koonce},
  \citenamefont {Cohen}, \citenamefont {Schooley}, \citenamefont {Hosler},\
  and\ \citenamefont {Pfeiffer}}]{Koonce:1967fe}%
  \BibitemOpen
  \bibfield  {author} {\bibinfo {author} {\bibfnamefont {C.~S.}\ \bibnamefont
  {Koonce}}, \bibinfo {author} {\bibfnamefont {M.~L.}\ \bibnamefont {Cohen}},
  \bibinfo {author} {\bibfnamefont {J.~F.}\ \bibnamefont {Schooley}}, \bibinfo
  {author} {\bibfnamefont {W.~R.}\ \bibnamefont {Hosler}},\ and\ \bibinfo
  {author} {\bibfnamefont {E.~R.}\ \bibnamefont {Pfeiffer}},\ }\href@noop {}
  {\bibfield  {journal} {\bibinfo  {journal} {Phys. Rev.}\ }\textbf {\bibinfo
  {volume} {163}},\ \bibinfo {pages} {380} (\bibinfo {year}
  {1967})}\BibitemShut {NoStop}%
\bibitem [{\citenamefont {Muller}\ and\ \citenamefont
  {Burkard}(1979)}]{Muller:1979je}%
  \BibitemOpen
  \bibfield  {author} {\bibinfo {author} {\bibfnamefont {K.~A.}\ \bibnamefont
  {Muller}}\ and\ \bibinfo {author} {\bibfnamefont {H.}~\bibnamefont
  {Burkard}},\ }\href@noop {} {\bibfield  {journal} {\bibinfo  {journal} {Phys.
  Rev. B}\ }\textbf {\bibinfo {volume} {19}},\ \bibinfo {pages} {3593}
  (\bibinfo {year} {1979})}\BibitemShut {NoStop}%
\bibitem [{\citenamefont {Haeni}\ \emph {et~al.}(2004)\citenamefont {Haeni},
  \citenamefont {Irvin}, \citenamefont {Chang}, \citenamefont {Uecker},
  \citenamefont {Reiche}, \citenamefont {Li}, \citenamefont {Choudhury},
  \citenamefont {Tian}, \citenamefont {Hawley}, \citenamefont {Craigo},
  \citenamefont {Tagantsev}, \citenamefont {Pan}, \citenamefont {Streiffer},
  \citenamefont {Chen}, \citenamefont {Kirchoefer}, \citenamefont {Levy},\ and\
  \citenamefont {Schlom}}]{Haeni2004}%
  \BibitemOpen
  \bibfield  {author} {\bibinfo {author} {\bibfnamefont {J.~H.}\ \bibnamefont
  {Haeni}}, \bibinfo {author} {\bibfnamefont {P.}~\bibnamefont {Irvin}},
  \bibinfo {author} {\bibfnamefont {W.}~\bibnamefont {Chang}}, \bibinfo
  {author} {\bibfnamefont {R.}~\bibnamefont {Uecker}}, \bibinfo {author}
  {\bibfnamefont {P.}~\bibnamefont {Reiche}}, \bibinfo {author} {\bibfnamefont
  {Y.~L.}\ \bibnamefont {Li}}, \bibinfo {author} {\bibfnamefont
  {S.}~\bibnamefont {Choudhury}}, \bibinfo {author} {\bibfnamefont
  {W.}~\bibnamefont {Tian}}, \bibinfo {author} {\bibfnamefont {M.~E.}\
  \bibnamefont {Hawley}}, \bibinfo {author} {\bibfnamefont {B.}~\bibnamefont
  {Craigo}}, \bibinfo {author} {\bibfnamefont {A.~K.}\ \bibnamefont
  {Tagantsev}}, \bibinfo {author} {\bibfnamefont {X.~Q.}\ \bibnamefont {Pan}},
  \bibinfo {author} {\bibfnamefont {S.~K.}\ \bibnamefont {Streiffer}}, \bibinfo
  {author} {\bibfnamefont {L.~Q.}\ \bibnamefont {Chen}}, \bibinfo {author}
  {\bibfnamefont {S.~W.}\ \bibnamefont {Kirchoefer}}, \bibinfo {author}
  {\bibfnamefont {J.}~\bibnamefont {Levy}},\ and\ \bibinfo {author}
  {\bibfnamefont {D.~G.}\ \bibnamefont {Schlom}},\ }\href@noop {} {\bibfield
  {journal} {\bibinfo  {journal} {Nature}\ }\textbf {\bibinfo {volume} {430}},\
  \bibinfo {pages} {758} (\bibinfo {year} {2004})}\BibitemShut {NoStop}%
\bibitem [{\citenamefont {Itoh}\ \emph {et~al.}(1999)\citenamefont {Itoh},
  \citenamefont {Wang}, \citenamefont {Inaguma}, \citenamefont {Yamaguchi},
  \citenamefont {Shan},\ and\ \citenamefont {Nakamura}}]{Itoh:1999eq}%
  \BibitemOpen
  \bibfield  {author} {\bibinfo {author} {\bibfnamefont {M.}~\bibnamefont
  {Itoh}}, \bibinfo {author} {\bibfnamefont {R.}~\bibnamefont {Wang}}, \bibinfo
  {author} {\bibfnamefont {Y.}~\bibnamefont {Inaguma}}, \bibinfo {author}
  {\bibfnamefont {T.}~\bibnamefont {Yamaguchi}}, \bibinfo {author}
  {\bibfnamefont {Y.-J.}\ \bibnamefont {Shan}},\ and\ \bibinfo {author}
  {\bibfnamefont {T.}~\bibnamefont {Nakamura}},\ }\href@noop {} {\bibfield
  {journal} {\bibinfo  {journal} {Phys. Rev. Lett.}\ }\textbf {\bibinfo
  {volume} {82}},\ \bibinfo {pages} {3540} (\bibinfo {year}
  {1999})}\BibitemShut {NoStop}%
\bibitem [{\citenamefont {Li}\ \emph {et~al.}(2019)\citenamefont {Li},
  \citenamefont {Qiu}, \citenamefont {Zhang}, \citenamefont {Baldini},
  \citenamefont {Lu}, \citenamefont {Rappe},\ and\ \citenamefont
  {Nelson}}]{Li2019}%
  \BibitemOpen
  \bibfield  {author} {\bibinfo {author} {\bibfnamefont {X.}~\bibnamefont
  {Li}}, \bibinfo {author} {\bibfnamefont {T.}~\bibnamefont {Qiu}}, \bibinfo
  {author} {\bibfnamefont {J.}~\bibnamefont {Zhang}}, \bibinfo {author}
  {\bibfnamefont {E.}~\bibnamefont {Baldini}}, \bibinfo {author} {\bibfnamefont
  {J.}~\bibnamefont {Lu}}, \bibinfo {author} {\bibfnamefont {A.~M.}\
  \bibnamefont {Rappe}},\ and\ \bibinfo {author} {\bibfnamefont {K.~A.}\
  \bibnamefont {Nelson}},\ }\href@noop {} {\bibfield  {journal} {\bibinfo
  {journal} {Science}\ }\textbf {\bibinfo {volume} {364}},\ \bibinfo {pages}
  {1079} (\bibinfo {year} {2019})}\BibitemShut {NoStop}%
\bibitem [{\citenamefont {Nova}\ \emph {et~al.}(2019)\citenamefont {Nova},
  \citenamefont {Disa}, \citenamefont {Fechner},\ and\ \citenamefont
  {Cavalleri}}]{Nova2019}%
  \BibitemOpen
  \bibfield  {author} {\bibinfo {author} {\bibfnamefont {T.~F.}\ \bibnamefont
  {Nova}}, \bibinfo {author} {\bibfnamefont {A.~S.}\ \bibnamefont {Disa}},
  \bibinfo {author} {\bibfnamefont {M.}~\bibnamefont {Fechner}},\ and\ \bibinfo
  {author} {\bibfnamefont {A.}~\bibnamefont {Cavalleri}},\ }\href@noop {}
  {\bibfield  {journal} {\bibinfo  {journal} {Science}\ }\textbf {\bibinfo
  {volume} {364}},\ \bibinfo {pages} {1075} (\bibinfo {year}
  {2019})}\BibitemShut {NoStop}%
\bibitem [{\citenamefont {Li}\ \emph {et~al.}(2006)\citenamefont {Li},
  \citenamefont {Choudhury}, \citenamefont {Haeni}, \citenamefont {Biegalski},
  \citenamefont {Vasudevarao}, \citenamefont {Sharan}, \citenamefont {Ma},
  \citenamefont {Levy}, \citenamefont {Gopalan}, \citenamefont
  {Trolier-McKinstry}, \citenamefont {Schlom}, \citenamefont {Jia},\ and\
  \citenamefont {Chen}}]{Li:2006}%
  \BibitemOpen
  \bibfield  {author} {\bibinfo {author} {\bibfnamefont {Y.~L.}\ \bibnamefont
  {Li}}, \bibinfo {author} {\bibfnamefont {S.}~\bibnamefont {Choudhury}},
  \bibinfo {author} {\bibfnamefont {J.~H.}\ \bibnamefont {Haeni}}, \bibinfo
  {author} {\bibfnamefont {M.~D.}\ \bibnamefont {Biegalski}}, \bibinfo {author}
  {\bibfnamefont {A.}~\bibnamefont {Vasudevarao}}, \bibinfo {author}
  {\bibfnamefont {A.}~\bibnamefont {Sharan}}, \bibinfo {author} {\bibfnamefont
  {H.~Z.}\ \bibnamefont {Ma}}, \bibinfo {author} {\bibfnamefont
  {J.}~\bibnamefont {Levy}}, \bibinfo {author} {\bibfnamefont {V.}~\bibnamefont
  {Gopalan}}, \bibinfo {author} {\bibfnamefont {S.}~\bibnamefont
  {Trolier-McKinstry}}, \bibinfo {author} {\bibfnamefont {D.~G.}\ \bibnamefont
  {Schlom}}, \bibinfo {author} {\bibfnamefont {Q.~X.}\ \bibnamefont {Jia}},\
  and\ \bibinfo {author} {\bibfnamefont {L.~Q.}\ \bibnamefont {Chen}},\
  }\href@noop {} {\bibfield  {journal} {\bibinfo  {journal} {Phys. Rev. B}\
  }\textbf {\bibinfo {volume} {73}},\ \bibinfo {pages} {184112} (\bibinfo
  {year} {2006})}\BibitemShut {NoStop}%
\bibitem [{\citenamefont {Appel}(1966)}]{Appel1966}%
  \BibitemOpen
  \bibfield  {author} {\bibinfo {author} {\bibfnamefont {J.}~\bibnamefont
  {Appel}},\ }\href@noop {} {\bibfield  {journal} {\bibinfo  {journal} {Phys.
  Rev. Lett}\ }\textbf {\bibinfo {volume} {17}},\ \bibinfo {pages} {1045}
  (\bibinfo {year} {1966})}\BibitemShut {NoStop}%
\bibitem [{\citenamefont {Rowley}\ \emph {et~al.}(2014)\citenamefont {Rowley},
  \citenamefont {Spalek}, \citenamefont {Smith}, \citenamefont {Dean},
  \citenamefont {Itoh}, \citenamefont {Scott}, \citenamefont {Lonzarich},\ and\
  \citenamefont {Saxena}}]{Rowley2014}%
  \BibitemOpen
  \bibfield  {author} {\bibinfo {author} {\bibfnamefont {S.~E.}\ \bibnamefont
  {Rowley}}, \bibinfo {author} {\bibfnamefont {L.~J.}\ \bibnamefont {Spalek}},
  \bibinfo {author} {\bibfnamefont {R.~P.}\ \bibnamefont {Smith}}, \bibinfo
  {author} {\bibfnamefont {M.~P.}\ \bibnamefont {Dean}}, \bibinfo {author}
  {\bibfnamefont {M.}~\bibnamefont {Itoh}}, \bibinfo {author} {\bibfnamefont
  {J.~F.}\ \bibnamefont {Scott}}, \bibinfo {author} {\bibfnamefont {G.~G.}\
  \bibnamefont {Lonzarich}},\ and\ \bibinfo {author} {\bibfnamefont {S.~S.}\
  \bibnamefont {Saxena}},\ }\href@noop {} {\bibfield  {journal} {\bibinfo
  {journal} {Nat. Phys.}\ }\textbf {\bibinfo {volume} {10}},\ \bibinfo {pages}
  {367} (\bibinfo {year} {2014})}\BibitemShut {NoStop}%
\bibitem [{\citenamefont {Edge}\ \emph {et~al.}(2015)\citenamefont {Edge},
  \citenamefont {Kedem}, \citenamefont {Aschauer}, \citenamefont {Spaldin},\
  and\ \citenamefont {Balatsky}}]{Edge:2015fj}%
  \BibitemOpen
  \bibfield  {author} {\bibinfo {author} {\bibfnamefont {J.~M.}\ \bibnamefont
  {Edge}}, \bibinfo {author} {\bibfnamefont {Y.}~\bibnamefont {Kedem}},
  \bibinfo {author} {\bibfnamefont {U.}~\bibnamefont {Aschauer}}, \bibinfo
  {author} {\bibfnamefont {N.~A.}\ \bibnamefont {Spaldin}},\ and\ \bibinfo
  {author} {\bibfnamefont {A.~V.}\ \bibnamefont {Balatsky}},\ }\href@noop {}
  {\bibfield  {journal} {\bibinfo  {journal} {Phys. Rev. Lett.}\ }\textbf
  {\bibinfo {volume} {115}},\ \bibinfo {pages} {247002} (\bibinfo {year}
  {2015})}\BibitemShut {NoStop}%
\bibitem [{\citenamefont {Rischau}\ \emph {et~al.}(2017)\citenamefont
  {Rischau}, \citenamefont {Lin}, \citenamefont {Grams}, \citenamefont {Finck},
  \citenamefont {Harms}, \citenamefont {Engelmayer}, \citenamefont {Lorenz},
  \citenamefont {Gallais}, \citenamefont {Fauque}, \citenamefont {Hemberger},\
  and\ \citenamefont {Behnia}}]{Rischau2017}%
  \BibitemOpen
  \bibfield  {author} {\bibinfo {author} {\bibfnamefont {C.~W.}\ \bibnamefont
  {Rischau}}, \bibinfo {author} {\bibfnamefont {X.}~\bibnamefont {Lin}},
  \bibinfo {author} {\bibfnamefont {C.~P.}\ \bibnamefont {Grams}}, \bibinfo
  {author} {\bibfnamefont {D.}~\bibnamefont {Finck}}, \bibinfo {author}
  {\bibfnamefont {S.}~\bibnamefont {Harms}}, \bibinfo {author} {\bibfnamefont
  {J.}~\bibnamefont {Engelmayer}}, \bibinfo {author} {\bibfnamefont
  {T.}~\bibnamefont {Lorenz}}, \bibinfo {author} {\bibfnamefont
  {Y.}~\bibnamefont {Gallais}}, \bibinfo {author} {\bibfnamefont
  {B.}~\bibnamefont {Fauque}}, \bibinfo {author} {\bibfnamefont
  {J.}~\bibnamefont {Hemberger}},\ and\ \bibinfo {author} {\bibfnamefont
  {K.}~\bibnamefont {Behnia}},\ }\href@noop {} {\bibfield  {journal} {\bibinfo
  {journal} {Nat. Phys.}\ }\textbf {\bibinfo {volume} {13}},\ \bibinfo {pages}
  {643} (\bibinfo {year} {2017})}\BibitemShut {NoStop}%
\bibitem [{\citenamefont {Narayan}\ \emph {et~al.}(2019)\citenamefont
  {Narayan}, \citenamefont {Cano}, \citenamefont {Balatsky},\ and\
  \citenamefont {Spaldin}}]{Narayan2019}%
  \BibitemOpen
  \bibfield  {author} {\bibinfo {author} {\bibfnamefont {A.}~\bibnamefont
  {Narayan}}, \bibinfo {author} {\bibfnamefont {A.}~\bibnamefont {Cano}},
  \bibinfo {author} {\bibfnamefont {A.~V.}\ \bibnamefont {Balatsky}},\ and\
  \bibinfo {author} {\bibfnamefont {N.~A.}\ \bibnamefont {Spaldin}},\
  }\href@noop {} {\bibfield  {journal} {\bibinfo  {journal} {Nat. Mater.}\
  }\textbf {\bibinfo {volume} {18}},\ \bibinfo {pages} {223} (\bibinfo {year}
  {2019})}\BibitemShut {NoStop}%
\bibitem [{\citenamefont {Itahashi}\ \emph {et~al.}(2020)\citenamefont
  {Itahashi}, \citenamefont {Ideue}, \citenamefont {Saito}, \citenamefont
  {Shimizu}, \citenamefont {Ouchi}, \citenamefont {Nojima},\ and\ \citenamefont
  {Iwasa}}]{Itahashi2020}%
  \BibitemOpen
  \bibfield  {author} {\bibinfo {author} {\bibfnamefont {Y.~M.}\ \bibnamefont
  {Itahashi}}, \bibinfo {author} {\bibfnamefont {T.}~\bibnamefont {Ideue}},
  \bibinfo {author} {\bibfnamefont {Y.}~\bibnamefont {Saito}}, \bibinfo
  {author} {\bibfnamefont {S.}~\bibnamefont {Shimizu}}, \bibinfo {author}
  {\bibfnamefont {T.}~\bibnamefont {Ouchi}}, \bibinfo {author} {\bibfnamefont
  {T.}~\bibnamefont {Nojima}},\ and\ \bibinfo {author} {\bibfnamefont
  {Y.}~\bibnamefont {Iwasa}},\ }\href@noop {} {\bibfield  {journal} {\bibinfo
  {journal} {Sci. Adv.}\ }\textbf {\bibinfo {volume} {6}},\ \bibinfo {pages}
  {eaay9120} (\bibinfo {year} {2020})}\BibitemShut {NoStop}%
\bibitem [{\citenamefont {Shirane}\ \emph {et~al.}(1952)\citenamefont
  {Shirane}, \citenamefont {Suzuki},\ and\ \citenamefont
  {Takeda}}]{Shirane1952}%
  \BibitemOpen
  \bibfield  {author} {\bibinfo {author} {\bibfnamefont {G.}~\bibnamefont
  {Shirane}}, \bibinfo {author} {\bibfnamefont {K.}~\bibnamefont {Suzuki}},\
  and\ \bibinfo {author} {\bibfnamefont {A.}~\bibnamefont {Takeda}},\
  }\href@noop {} {\bibfield  {journal} {\bibinfo  {journal} {J. Phys. Soc.
  Japan}\ }\textbf {\bibinfo {volume} {7}},\ \bibinfo {pages} {12} (\bibinfo
  {year} {1952})}\BibitemShut {NoStop}%
\bibitem [{\citenamefont {Samara}(1971)}]{Samara1971}%
  \BibitemOpen
  \bibfield  {author} {\bibinfo {author} {\bibfnamefont {G.~A.}\ \bibnamefont
  {Samara}},\ }\href@noop {} {\bibfield  {journal} {\bibinfo  {journal}
  {Ferroelectrics}\ }\textbf {\bibinfo {volume} {2}},\ \bibinfo {pages} {277}
  (\bibinfo {year} {1971})}\BibitemShut {NoStop}%
\bibitem [{\citenamefont {Zhong}\ and\ \citenamefont
  {Vanderbilt}(1995)}]{Zhong1995}%
  \BibitemOpen
  \bibfield  {author} {\bibinfo {author} {\bibfnamefont {W.}~\bibnamefont
  {Zhong}}\ and\ \bibinfo {author} {\bibfnamefont {D.}~\bibnamefont
  {Vanderbilt}},\ }\href@noop {} {\bibfield  {journal} {\bibinfo  {journal}
  {Phys. Rev. Lett.}\ }\textbf {\bibinfo {volume} {74}},\ \bibinfo {pages}
  {2587} (\bibinfo {year} {1995})}\BibitemShut {NoStop}%
\bibitem [{\citenamefont {Bellaiche}\ \emph {et~al.}(2000)\citenamefont
  {Bellaiche}, \citenamefont {Garcia},\ and\ \citenamefont
  {Vanderbilt}}]{Bellaiche2000}%
  \BibitemOpen
  \bibfield  {author} {\bibinfo {author} {\bibfnamefont {L.}~\bibnamefont
  {Bellaiche}}, \bibinfo {author} {\bibfnamefont {A.}~\bibnamefont {Garcia}},\
  and\ \bibinfo {author} {\bibfnamefont {D.}~\bibnamefont {Vanderbilt}},\
  }\href@noop {} {\bibfield  {journal} {\bibinfo  {journal} {Phys. Rev. Lett.}\
  }\textbf {\bibinfo {volume} {84}},\ \bibinfo {pages} {5427} (\bibinfo {year}
  {2000})}\BibitemShut {NoStop}%
\bibitem [{\citenamefont {Guo}\ \emph {et~al.}(2000)\citenamefont {Guo},
  \citenamefont {Cross}, \citenamefont {Park}, \citenamefont {Noheda},
  \citenamefont {Cox},\ and\ \citenamefont {Shirane}}]{Guo2000}%
  \BibitemOpen
  \bibfield  {author} {\bibinfo {author} {\bibfnamefont {R.}~\bibnamefont
  {Guo}}, \bibinfo {author} {\bibfnamefont {L.~E.}\ \bibnamefont {Cross}},
  \bibinfo {author} {\bibfnamefont {S.~E.}\ \bibnamefont {Park}}, \bibinfo
  {author} {\bibfnamefont {B.}~\bibnamefont {Noheda}}, \bibinfo {author}
  {\bibfnamefont {D.~E.}\ \bibnamefont {Cox}},\ and\ \bibinfo {author}
  {\bibfnamefont {G.}~\bibnamefont {Shirane}},\ }\href@noop {} {\bibfield
  {journal} {\bibinfo  {journal} {Phys. Rev. Lett.}\ }\textbf {\bibinfo
  {volume} {84}},\ \bibinfo {pages} {5423} (\bibinfo {year}
  {2000})}\BibitemShut {NoStop}%
\bibitem [{\citenamefont {Miyasaka}\ \emph {et~al.}(2003)\citenamefont
  {Miyasaka}, \citenamefont {Okimoto}, \citenamefont {Iwama},\ and\
  \citenamefont {Tokura}}]{Miyasaka2003}%
  \BibitemOpen
  \bibfield  {author} {\bibinfo {author} {\bibfnamefont {S.}~\bibnamefont
  {Miyasaka}}, \bibinfo {author} {\bibfnamefont {Y.}~\bibnamefont {Okimoto}},
  \bibinfo {author} {\bibfnamefont {M.}~\bibnamefont {Iwama}},\ and\ \bibinfo
  {author} {\bibfnamefont {Y.}~\bibnamefont {Tokura}},\ }\href@noop {}
  {\bibfield  {journal} {\bibinfo  {journal} {Phys. Rev. B}\ }\textbf {\bibinfo
  {volume} {68}},\ \bibinfo {pages} {100406(R)} (\bibinfo {year}
  {2003})}\BibitemShut {NoStop}%
\bibitem [{\citenamefont {Fennie}\ and\ \citenamefont
  {Rabe}(2006)}]{Fennie2006}%
  \BibitemOpen
  \bibfield  {author} {\bibinfo {author} {\bibfnamefont {C.~J.}\ \bibnamefont
  {Fennie}}\ and\ \bibinfo {author} {\bibfnamefont {K.~M.}\ \bibnamefont
  {Rabe}},\ }\href@noop {} {\bibfield  {journal} {\bibinfo  {journal} {Phys.
  Rev. Lett.}\ }\textbf {\bibinfo {volume} {97}},\ \bibinfo {pages} {267602}
  (\bibinfo {year} {2006})}\BibitemShut {NoStop}%
\bibitem [{\citenamefont {Lee}\ and\ \citenamefont {Rabe}(2010)}]{Lee2010}%
  \BibitemOpen
  \bibfield  {author} {\bibinfo {author} {\bibfnamefont {J.~H.}\ \bibnamefont
  {Lee}}\ and\ \bibinfo {author} {\bibfnamefont {K.~M.}\ \bibnamefont {Rabe}},\
  }\href@noop {} {\bibfield  {journal} {\bibinfo  {journal} {Phys. Rev. Lett.}\
  }\textbf {\bibinfo {volume} {104}},\ \bibinfo {pages} {207204} (\bibinfo
  {year} {2010})}\BibitemShut {NoStop}%
\bibitem [{\citenamefont {Rytz}\ \emph {et~al.}(1980)\citenamefont {Rytz},
  \citenamefont {Hochli},\ and\ \citenamefont {Bilz}}]{Rytz1980}%
  \BibitemOpen
  \bibfield  {author} {\bibinfo {author} {\bibfnamefont {D.}~\bibnamefont
  {Rytz}}, \bibinfo {author} {\bibfnamefont {U.~T.}\ \bibnamefont {Hochli}},\
  and\ \bibinfo {author} {\bibfnamefont {H.}~\bibnamefont {Bilz}},\ }\href@noop
  {} {\bibfield  {journal} {\bibinfo  {journal} {Phys. Rev. B}\ }\textbf
  {\bibinfo {volume} {22}},\ \bibinfo {pages} {359} (\bibinfo {year}
  {1980})}\BibitemShut {NoStop}%
\bibitem [{\citenamefont {Akbarzadeh}\ \emph {et~al.}(2004)\citenamefont
  {Akbarzadeh}, \citenamefont {Bellaiche}, \citenamefont {Leung}, \citenamefont
  {Iniguez},\ and\ \citenamefont {Vanderbilt}}]{Akbarzadeh2004}%
  \BibitemOpen
  \bibfield  {author} {\bibinfo {author} {\bibfnamefont {A.~R.}\ \bibnamefont
  {Akbarzadeh}}, \bibinfo {author} {\bibfnamefont {L.}~\bibnamefont
  {Bellaiche}}, \bibinfo {author} {\bibfnamefont {K.}~\bibnamefont {Leung}},
  \bibinfo {author} {\bibfnamefont {J.}~\bibnamefont {Iniguez}},\ and\ \bibinfo
  {author} {\bibfnamefont {D.}~\bibnamefont {Vanderbilt}},\ }\href@noop {}
  {\bibfield  {journal} {\bibinfo  {journal} {Phys. Rev. B}\ }\textbf {\bibinfo
  {volume} {70}},\ \bibinfo {pages} {054103} (\bibinfo {year}
  {2004})}\BibitemShut {NoStop}%
\bibitem [{\citenamefont {Zhang}\ \emph {et~al.}(2020)\citenamefont {Zhang},
  \citenamefont {Ye}, \citenamefont {Xiang},\ and\ \citenamefont
  {Li}}]{Zhang2020}%
  \BibitemOpen
  \bibfield  {author} {\bibinfo {author} {\bibfnamefont {X.}~\bibnamefont
  {Zhang}}, \bibinfo {author} {\bibfnamefont {Q.~J.}\ \bibnamefont {Ye}},
  \bibinfo {author} {\bibfnamefont {H.}~\bibnamefont {Xiang}},\ and\ \bibinfo
  {author} {\bibfnamefont {X.~Z.}\ \bibnamefont {Li}},\ }\href@noop {}
  {\bibfield  {journal} {\bibinfo  {journal} {Phys. Rev. B}\ }\textbf {\bibinfo
  {volume} {101}},\ \bibinfo {pages} {104102} (\bibinfo {year}
  {2020})}\BibitemShut {NoStop}%
\bibitem [{\citenamefont {Cowley}(1964)}]{Cowley1964}%
  \BibitemOpen
  \bibfield  {author} {\bibinfo {author} {\bibfnamefont {R.~A.}\ \bibnamefont
  {Cowley}},\ }\href@noop {} {\bibfield  {journal} {\bibinfo  {journal} {Phys.
  Rev.}\ }\textbf {\bibinfo {volume} {134}},\ \bibinfo {pages} {A981} (\bibinfo
  {year} {1964})}\BibitemShut {NoStop}%
\bibitem [{\citenamefont {Prosandeev}\ \emph {et~al.}(1999)\citenamefont
  {Prosandeev}, \citenamefont {Kleemann}, \citenamefont {Westwanski},\ and\
  \citenamefont {Dec}}]{Westwanski1999}%
  \BibitemOpen
  \bibfield  {author} {\bibinfo {author} {\bibfnamefont {S.~A.}\ \bibnamefont
  {Prosandeev}}, \bibinfo {author} {\bibfnamefont {W.}~\bibnamefont
  {Kleemann}}, \bibinfo {author} {\bibfnamefont {B.}~\bibnamefont
  {Westwanski}},\ and\ \bibinfo {author} {\bibfnamefont {J.}~\bibnamefont
  {Dec}},\ }\href@noop {} {\bibfield  {journal} {\bibinfo  {journal} {Phys.
  Rev. B}\ }\textbf {\bibinfo {volume} {60}},\ \bibinfo {pages} {14489}
  (\bibinfo {year} {1999})}\BibitemShut {NoStop}%
\bibitem [{\citenamefont {Marques}\ \emph {et~al.}(2005)\citenamefont
  {Marques}, \citenamefont {Arago},\ and\ \citenamefont
  {Gonzalo}}]{Marques2005}%
  \BibitemOpen
  \bibfield  {author} {\bibinfo {author} {\bibfnamefont {M.~I.}\ \bibnamefont
  {Marques}}, \bibinfo {author} {\bibfnamefont {C.}~\bibnamefont {Arago}},\
  and\ \bibinfo {author} {\bibfnamefont {J.~A.}\ \bibnamefont {Gonzalo}},\
  }\href@noop {} {\bibfield  {journal} {\bibinfo  {journal} {Phys. Rev. B}\
  }\textbf {\bibinfo {volume} {72}},\ \bibinfo {pages} {092103} (\bibinfo
  {year} {2005})}\BibitemShut {NoStop}%
\bibitem [{\citenamefont {Fujishita}\ \emph {et~al.}(2016)\citenamefont
  {Fujishita}, \citenamefont {Kitazawa}, \citenamefont {Saito}, \citenamefont
  {Ishisaka}, \citenamefont {Okamoto},\ and\ \citenamefont
  {Yamaguchi}}]{Fujishita2016}%
  \BibitemOpen
  \bibfield  {author} {\bibinfo {author} {\bibfnamefont {H.}~\bibnamefont
  {Fujishita}}, \bibinfo {author} {\bibfnamefont {S.}~\bibnamefont {Kitazawa}},
  \bibinfo {author} {\bibfnamefont {M.}~\bibnamefont {Saito}}, \bibinfo
  {author} {\bibfnamefont {R.}~\bibnamefont {Ishisaka}}, \bibinfo {author}
  {\bibfnamefont {H.}~\bibnamefont {Okamoto}},\ and\ \bibinfo {author}
  {\bibfnamefont {T.}~\bibnamefont {Yamaguchi}},\ }\href@noop {} {\bibfield
  {journal} {\bibinfo  {journal} {J. Phys. Soc. Japan}\ }\textbf {\bibinfo
  {volume} {85}},\ \bibinfo {pages} {074703} (\bibinfo {year}
  {2016})}\BibitemShut {NoStop}%
\bibitem [{\citenamefont {Zhong}\ and\ \citenamefont
  {Vanderbilt}(1996)}]{zhong1996}%
  \BibitemOpen
  \bibfield  {author} {\bibinfo {author} {\bibfnamefont {W.}~\bibnamefont
  {Zhong}}\ and\ \bibinfo {author} {\bibfnamefont {D.}~\bibnamefont
  {Vanderbilt}},\ }\href@noop {} {\bibfield  {journal} {\bibinfo  {journal}
  {Phys. Rev. B}\ }\textbf {\bibinfo {volume} {53}},\ \bibinfo {pages} {5047}
  (\bibinfo {year} {1996})}\BibitemShut {NoStop}%
\bibitem [{\citenamefont {Martonak}\ and\ \citenamefont
  {Tosatti}(1996)}]{Martonak1996}%
  \BibitemOpen
  \bibfield  {author} {\bibinfo {author} {\bibfnamefont {R.}~\bibnamefont
  {Martonak}}\ and\ \bibinfo {author} {\bibfnamefont {E.}~\bibnamefont
  {Tosatti}},\ }\href@noop {} {\bibfield  {journal} {\bibinfo  {journal} {Phys.
  Rev. B}\ }\textbf {\bibinfo {volume} {54}},\ \bibinfo {pages} {15714}
  (\bibinfo {year} {1996})}\BibitemShut {NoStop}%
\bibitem [{\citenamefont {Song}\ \emph {et~al.}(1996)\citenamefont {Song},
  \citenamefont {Kim}, \citenamefont {Kwun}, \citenamefont {Kim},\ and\
  \citenamefont {Kim}}]{Song1996}%
  \BibitemOpen
  \bibfield  {author} {\bibinfo {author} {\bibfnamefont {T.~K.}\ \bibnamefont
  {Song}}, \bibinfo {author} {\bibfnamefont {J.}~\bibnamefont {Kim}}, \bibinfo
  {author} {\bibfnamefont {S.~I.}\ \bibnamefont {Kwun}}, \bibinfo {author}
  {\bibfnamefont {C.~J.}\ \bibnamefont {Kim}},\ and\ \bibinfo {author}
  {\bibfnamefont {J.~J.}\ \bibnamefont {Kim}},\ }\href@noop {} {\bibfield
  {journal} {\bibinfo  {journal} {Phys. B Condens. Matter}\ }\textbf {\bibinfo
  {volume} {219-220}},\ \bibinfo {pages} {538} (\bibinfo {year}
  {1996})}\BibitemShut {NoStop}%
\bibitem [{\citenamefont {He}\ \emph {et~al.}(2020)\citenamefont {He},
  \citenamefont {Bansal}, \citenamefont {Winn}, \citenamefont {Chi},
  \citenamefont {Boatner},\ and\ \citenamefont {Delaire}}]{He2020}%
  \BibitemOpen
  \bibfield  {author} {\bibinfo {author} {\bibfnamefont {X.}~\bibnamefont
  {He}}, \bibinfo {author} {\bibfnamefont {D.}~\bibnamefont {Bansal}}, \bibinfo
  {author} {\bibfnamefont {B.}~\bibnamefont {Winn}}, \bibinfo {author}
  {\bibfnamefont {S.}~\bibnamefont {Chi}}, \bibinfo {author} {\bibfnamefont
  {L.}~\bibnamefont {Boatner}},\ and\ \bibinfo {author} {\bibfnamefont
  {O.}~\bibnamefont {Delaire}},\ }\href@noop {} {\bibfield  {journal} {\bibinfo
   {journal} {Phys. Rev. Lett.}\ }\textbf {\bibinfo {volume} {124}},\ \bibinfo
  {pages} {145901} (\bibinfo {year} {2020})}\BibitemShut {NoStop}%
\bibitem [{\citenamefont {Gonze}\ and\ \citenamefont {Lee}(1997)}]{Gonze1997}%
  \BibitemOpen
  \bibfield  {author} {\bibinfo {author} {\bibfnamefont {X.}~\bibnamefont
  {Gonze}}\ and\ \bibinfo {author} {\bibfnamefont {C.}~\bibnamefont {Lee}},\
  }\href@noop {} {\bibfield  {journal} {\bibinfo  {journal} {Phys. Rev. B}\
  }\textbf {\bibinfo {volume} {55}},\ \bibinfo {pages} {10355} (\bibinfo {year}
  {1997})}\BibitemShut {NoStop}%
\bibitem [{\citenamefont {Aschauer}\ and\ \citenamefont
  {Spaldin}(2014)}]{Aschauer:2014}%
  \BibitemOpen
  \bibfield  {author} {\bibinfo {author} {\bibfnamefont {U.}~\bibnamefont
  {Aschauer}}\ and\ \bibinfo {author} {\bibfnamefont {N.~A.}\ \bibnamefont
  {Spaldin}},\ }\href@noop {} {\bibfield  {journal} {\bibinfo  {journal} {J.
  Phys. Condens. Matter}\ }\textbf {\bibinfo {volume} {26}},\ \bibinfo {pages}
  {122203} (\bibinfo {year} {2014})}\BibitemShut {NoStop}%
\bibitem [{\citenamefont {Zhou}\ \emph {et~al.}(2018)\citenamefont {Zhou},
  \citenamefont {Hellman},\ and\ \citenamefont {Bernardi}}]{Zhou:2018}%
  \BibitemOpen
  \bibfield  {author} {\bibinfo {author} {\bibfnamefont {J.~J.}\ \bibnamefont
  {Zhou}}, \bibinfo {author} {\bibfnamefont {O.}~\bibnamefont {Hellman}},\ and\
  \bibinfo {author} {\bibfnamefont {M.}~\bibnamefont {Bernardi}},\ }\href@noop
  {} {\bibfield  {journal} {\bibinfo  {journal} {Phys. Rev. Lett.}\ }\textbf
  {\bibinfo {volume} {121}},\ \bibinfo {pages} {226603} (\bibinfo {year}
  {2018})}\BibitemShut {NoStop}%
\bibitem [{\citenamefont {Wahl}\ \emph {et~al.}(2008)\citenamefont {Wahl},
  \citenamefont {Vogtenhuber},\ and\ \citenamefont {Kresse}}]{Wahl:2008}%
  \BibitemOpen
  \bibfield  {author} {\bibinfo {author} {\bibfnamefont {R.}~\bibnamefont
  {Wahl}}, \bibinfo {author} {\bibfnamefont {D.}~\bibnamefont {Vogtenhuber}},\
  and\ \bibinfo {author} {\bibfnamefont {G.}~\bibnamefont {Kresse}},\
  }\href@noop {} {\bibfield  {journal} {\bibinfo  {journal} {Phys. Rev. B}\
  }\textbf {\bibinfo {volume} {78}},\ \bibinfo {pages} {104116} (\bibinfo
  {year} {2008})}\BibitemShut {NoStop}%
\bibitem [{\citenamefont {Cao}\ \emph {et~al.}(2000)\citenamefont {Cao},
  \citenamefont {Sozontov},\ and\ \citenamefont {Zecenhagen}}]{Cao2000}%
  \BibitemOpen
  \bibfield  {author} {\bibinfo {author} {\bibfnamefont {L.}~\bibnamefont
  {Cao}}, \bibinfo {author} {\bibfnamefont {E.}~\bibnamefont {Sozontov}},\ and\
  \bibinfo {author} {\bibfnamefont {J.}~\bibnamefont {Zecenhagen}},\
  }\href@noop {} {\bibfield  {journal} {\bibinfo  {journal} {Phys. Status
  Solidi Appl. Res.}\ }\textbf {\bibinfo {volume} {181}},\ \bibinfo {pages}
  {387} (\bibinfo {year} {2000})}\BibitemShut {NoStop}%
\bibitem [{\citenamefont {Heidemann}\ and\ \citenamefont
  {Wettengel}(1973)}]{Heidemann1973}%
  \BibitemOpen
  \bibfield  {author} {\bibinfo {author} {\bibfnamefont {A.}~\bibnamefont
  {Heidemann}}\ and\ \bibinfo {author} {\bibfnamefont {H.}~\bibnamefont
  {Wettengel}},\ }\href@noop {} {\bibfield  {journal} {\bibinfo  {journal}
  {Zeitschrift fur Phys.}\ }\textbf {\bibinfo {volume} {258}},\ \bibinfo
  {pages} {429} (\bibinfo {year} {1973})}\BibitemShut {NoStop}%
\bibitem [{\citenamefont {Unoki}\ and\ \citenamefont
  {Sakudo}(1967)}]{Unoki1967}%
  \BibitemOpen
  \bibfield  {author} {\bibinfo {author} {\bibfnamefont {H.}~\bibnamefont
  {Unoki}}\ and\ \bibinfo {author} {\bibfnamefont {T.}~\bibnamefont {Sakudo}},\
  }\href@noop {} {\bibfield  {journal} {\bibinfo  {journal} {J. Phys. Soc.
  Japan}\ }\textbf {\bibinfo {volume} {23}},\ \bibinfo {pages} {546} (\bibinfo
  {year} {1967})}\BibitemShut {NoStop}%
\bibitem [{\citenamefont {Giannozzi}\ \emph {et~al.}(2017)\citenamefont
  {Giannozzi} \emph {et~al.}}]{Giannozzi2017}%
  \BibitemOpen
  \bibfield  {author} {\bibinfo {author} {\bibfnamefont {P.}~\bibnamefont
  {Giannozzi}} \emph {et~al.},\ }\href@noop {} {\bibfield  {journal} {\bibinfo
  {journal} {J. Condens. Matter Phys.}\ }\textbf {\bibinfo {volume} {29}},\
  \bibinfo {pages} {465901} (\bibinfo {year} {2017})}\BibitemShut {NoStop}%
\bibitem [{\citenamefont {Loetzsch}\ \emph {et~al.}(2010)\citenamefont
  {Loetzsch}, \citenamefont {Lubcke}, \citenamefont {Uschmann}, \citenamefont
  {Forster}, \citenamefont {Groe}, \citenamefont {Thuerk}, \citenamefont
  {Koettig}, \citenamefont {Schmidl},\ and\ \citenamefont
  {Seidel}}]{Loetzsch2010}%
  \BibitemOpen
  \bibfield  {author} {\bibinfo {author} {\bibfnamefont {R.}~\bibnamefont
  {Loetzsch}}, \bibinfo {author} {\bibfnamefont {A.}~\bibnamefont {Lubcke}},
  \bibinfo {author} {\bibfnamefont {I.}~\bibnamefont {Uschmann}}, \bibinfo
  {author} {\bibfnamefont {E.}~\bibnamefont {Forster}}, \bibinfo {author}
  {\bibfnamefont {V.}~\bibnamefont {Groe}}, \bibinfo {author} {\bibfnamefont
  {M.}~\bibnamefont {Thuerk}}, \bibinfo {author} {\bibfnamefont
  {T.}~\bibnamefont {Koettig}}, \bibinfo {author} {\bibfnamefont
  {F.}~\bibnamefont {Schmidl}},\ and\ \bibinfo {author} {\bibfnamefont
  {P.}~\bibnamefont {Seidel}},\ }\href@noop {} {\bibfield  {journal} {\bibinfo
  {journal} {Appl. Phys. Lett.}\ }\textbf {\bibinfo {volume} {96}},\ \bibinfo
  {pages} {071901} (\bibinfo {year} {2010})}\BibitemShut {NoStop}%
\bibitem [{\citenamefont {Perdew}\ and\ \citenamefont
  {Wang}(1992)}]{Perdew1992}%
  \BibitemOpen
  \bibfield  {author} {\bibinfo {author} {\bibfnamefont {J.~P.}\ \bibnamefont
  {Perdew}}\ and\ \bibinfo {author} {\bibfnamefont {Y.}~\bibnamefont {Wang}},\
  }\href@noop {} {\bibfield  {journal} {\bibinfo  {journal} {Phys. Rev. B}\
  }\textbf {\bibinfo {volume} {45}},\ \bibinfo {pages} {13244} (\bibinfo {year}
  {1992})}\BibitemShut {NoStop}%
\bibitem [{\citenamefont {Csonka}\ \emph {et~al.}(2009)\citenamefont {Csonka},
  \citenamefont {Perdew}, \citenamefont {Ruzsinszky}, \citenamefont
  {Philipsen}, \citenamefont {Lebegue}, \citenamefont {Paier}, \citenamefont
  {Vydrov},\ and\ \citenamefont {Angyan}}]{Csonka2009}%
  \BibitemOpen
  \bibfield  {author} {\bibinfo {author} {\bibfnamefont {G.~I.}\ \bibnamefont
  {Csonka}}, \bibinfo {author} {\bibfnamefont {J.~P.}\ \bibnamefont {Perdew}},
  \bibinfo {author} {\bibfnamefont {A.}~\bibnamefont {Ruzsinszky}}, \bibinfo
  {author} {\bibfnamefont {P.H.T.}~\bibnamefont {Philipsen}}, \bibinfo {author}
  {\bibfnamefont {S.}~\bibnamefont {Lebegue}}, \bibinfo {author} {\bibfnamefont
  {J.}~\bibnamefont {Paier}}, \bibinfo {author} {\bibfnamefont {O.~A.}\
  \bibnamefont {Vydrov}},\ and\ \bibinfo {author} {\bibfnamefont {J.~G.}\
  \bibnamefont {Angyan}},\ }\href@noop {} {\bibfield  {journal} {\bibinfo
  {journal} {Phys. Rev. B}\ }\textbf {\bibinfo {volume} {79}},\ \bibinfo
  {pages} {155107} (\bibinfo {year} {2009})}\BibitemShut {NoStop}%
\bibitem [{\citenamefont {Perdew}\ \emph {et~al.}(1996)\citenamefont {Perdew},
  \citenamefont {Burke},\ and\ \citenamefont {Ernzerhof}}]{Perdew1996}%
  \BibitemOpen
  \bibfield  {author} {\bibinfo {author} {\bibfnamefont {J.~P.}\ \bibnamefont
  {Perdew}}, \bibinfo {author} {\bibfnamefont {K.}~\bibnamefont {Burke}},\ and\
  \bibinfo {author} {\bibfnamefont {M.}~\bibnamefont {Ernzerhof}},\ }\href@noop
  {} {\bibfield  {journal} {\bibinfo  {journal} {Phys. Rev. Lett.}\ }\textbf
  {\bibinfo {volume} {77}},\ \bibinfo {pages} {3865} (\bibinfo {year}
  {1996})}\BibitemShut {NoStop}%
\bibitem [{\citenamefont {Heyd}\ \emph {et~al.}(2003)\citenamefont {Heyd},
  \citenamefont {Scuseria},\ and\ \citenamefont {Ernzerhof}}]{Heyd2003}%
  \BibitemOpen
  \bibfield  {author} {\bibinfo {author} {\bibfnamefont {J.}~\bibnamefont
  {Heyd}}, \bibinfo {author} {\bibfnamefont {G.~E.}\ \bibnamefont {Scuseria}},\
  and\ \bibinfo {author} {\bibfnamefont {M.}~\bibnamefont {Ernzerhof}},\
  }\href@noop {} {\bibfield  {journal} {\bibinfo  {journal} {J. Chem. Phys.}\
  }\textbf {\bibinfo {volume} {118}},\ \bibinfo {pages} {8207} (\bibinfo {year}
  {2003})}\BibitemShut {NoStop}%
\bibitem [{\citenamefont {Krukau}\ \emph {et~al.}(2006)\citenamefont {Krukau},
  \citenamefont {Vydrov}, \citenamefont {Izmaylov},\ and\ \citenamefont
  {Scuseria}}]{Krukau2006}%
  \BibitemOpen
  \bibfield  {author} {\bibinfo {author} {\bibfnamefont {A.~V.}\ \bibnamefont
  {Krukau}}, \bibinfo {author} {\bibfnamefont {O.~A.}\ \bibnamefont {Vydrov}},
  \bibinfo {author} {\bibfnamefont {A.~F.}\ \bibnamefont {Izmaylov}},\ and\
  \bibinfo {author} {\bibfnamefont {G.~E.}\ \bibnamefont {Scuseria}},\
  }\href@noop {} {\bibfield  {journal} {\bibinfo  {journal} {J. Chem. Phys.}\
  }\textbf {\bibinfo {volume} {125}},\ \bibinfo {pages} {224106} (\bibinfo
  {year} {2006})}\BibitemShut {NoStop}%
\bibitem [{\citenamefont {El-Mellouhi}\ \emph {et~al.}(2011)\citenamefont
  {El-Mellouhi}, \citenamefont {Brothers}, \citenamefont {Lucero},\ and\
  \citenamefont {Scuseria}}]{El-Mellouhi2011}%
  \BibitemOpen
  \bibfield  {author} {\bibinfo {author} {\bibfnamefont {F.}~\bibnamefont
  {El-Mellouhi}}, \bibinfo {author} {\bibfnamefont {E.~N.}\ \bibnamefont
  {Brothers}}, \bibinfo {author} {\bibfnamefont {M.~J.}\ \bibnamefont
  {Lucero}},\ and\ \bibinfo {author} {\bibfnamefont {G.~E.}\ \bibnamefont
  {Scuseria}},\ }\href@noop {} {\bibfield  {journal} {\bibinfo  {journal}
  {Phys. Rev. B}\ }\textbf {\bibinfo {volume} {84}},\ \bibinfo {pages} {115122}
  (\bibinfo {year} {2011})}\BibitemShut {NoStop}%
\bibitem [{\citenamefont {Antons}\ \emph {et~al.}(2005)\citenamefont {Antons},
  \citenamefont {Neaton}, \citenamefont {Rabe},\ and\ \citenamefont
  {Vanderbilt}}]{Antons2005}%
  \BibitemOpen
  \bibfield  {author} {\bibinfo {author} {\bibfnamefont {A.}~\bibnamefont
  {Antons}}, \bibinfo {author} {\bibfnamefont {J.~B.}\ \bibnamefont {Neaton}},
  \bibinfo {author} {\bibfnamefont {K.~M.}\ \bibnamefont {Rabe}},\ and\
  \bibinfo {author} {\bibfnamefont {D.}~\bibnamefont {Vanderbilt}},\
  }\href@noop {} {\bibfield  {journal} {\bibinfo  {journal} {Phys. Rev. B}\
  }\textbf {\bibinfo {volume} {71}},\ \bibinfo {pages} {024102} (\bibinfo
  {year} {2005})}\BibitemShut {NoStop}%
\bibitem [{\citenamefont {Shirane}\ and\ \citenamefont
  {Yamada}(1969)}]{Shirane1969}%
  \BibitemOpen
  \bibfield  {author} {\bibinfo {author} {\bibfnamefont {G.}~\bibnamefont
  {Shirane}}\ and\ \bibinfo {author} {\bibfnamefont {Y.}~\bibnamefont
  {Yamada}},\ }\href@noop {} {\bibfield  {journal} {\bibinfo  {journal} {Phys.
  Rev.}\ }\textbf {\bibinfo {volume} {177}},\ \bibinfo {pages} {858} (\bibinfo
  {year} {1969})}\BibitemShut {NoStop}%
\bibitem [{\citenamefont {Yamanaka}\ \emph {et~al.}(2000)\citenamefont
  {Yamanaka}, \citenamefont {Kataoka}, \citenamefont {Inaba}, \citenamefont
  {Inoue}, \citenamefont {Hehlen},\ and\ \citenamefont
  {Courtens}}]{Yamanaka2000}%
  \BibitemOpen
  \bibfield  {author} {\bibinfo {author} {\bibfnamefont {A.}~\bibnamefont
  {Yamanaka}}, \bibinfo {author} {\bibfnamefont {M.}~\bibnamefont {Kataoka}},
  \bibinfo {author} {\bibfnamefont {Y.}~\bibnamefont {Inaba}}, \bibinfo
  {author} {\bibfnamefont {K.}~\bibnamefont {Inoue}}, \bibinfo {author}
  {\bibfnamefont {B.}~\bibnamefont {Hehlen}},\ and\ \bibinfo {author}
  {\bibfnamefont {E.}~\bibnamefont {Courtens}},\ }\href@noop {} {\bibfield
  {journal} {\bibinfo  {journal} {Europhys. Lett.}\ }\textbf {\bibinfo {volume}
  {50}},\ \bibinfo {pages} {688} (\bibinfo {year} {2000})}\BibitemShut
  {NoStop}%
\bibitem [{\citenamefont {Vogt}(1995)}]{Vogt1995}%
  \BibitemOpen
  \bibfield  {author} {\bibinfo {author} {\bibfnamefont {H.}~\bibnamefont
  {Vogt}},\ }\href@noop {} {\bibfield  {journal} {\bibinfo  {journal} {Phys.
  Rev. B}\ }\textbf {\bibinfo {volume} {51}},\ \bibinfo {pages} {8046}
  (\bibinfo {year} {1995})}\BibitemShut {NoStop}%
\bibitem [{SM()}]{SM}%
  \BibitemOpen
  \href@noop {} {\bibinfo {title} {See supplemental material at
  url}}\BibitemShut {NoStop}%
\bibitem [{\citenamefont {Wentzcovitch}(1991)}]{Wentzcovitch1991}%
  \BibitemOpen
  \bibfield  {author} {\bibinfo {author} {\bibfnamefont {R.~M.}\ \bibnamefont
  {Wentzcovitch}},\ }\href@noop {} {\bibfield  {journal} {\bibinfo  {journal}
  {Phys. Rev. B}\ }\textbf {\bibinfo {volume} {44}},\ \bibinfo {pages} {2358}
  (\bibinfo {year} {1991})}\BibitemShut {NoStop}%
\bibitem [{\citenamefont {Lyddane}\ \emph {et~al.}(1941)\citenamefont
  {Lyddane}, \citenamefont {Sachs},\ and\ \citenamefont
  {Teller}}]{Lyddane1941}%
  \BibitemOpen
  \bibfield  {author} {\bibinfo {author} {\bibfnamefont {R.}~\bibnamefont
  {Lyddane}}, \bibinfo {author} {\bibfnamefont {R.~G.}\ \bibnamefont {Sachs}},\
  and\ \bibinfo {author} {\bibfnamefont {E.}~\bibnamefont {Teller}},\
  }\href@noop {} {\bibfield  {journal} {\bibinfo  {journal} {Phys. Rev.}\
  }\textbf {\bibinfo {volume} {59}},\ \bibinfo {pages} {673} (\bibinfo {year}
  {1941})}\BibitemShut {NoStop}%
\bibitem [{\citenamefont {Worlock}\ and\ \citenamefont
  {Fleury}(1967)}]{Worlock1967}%
  \BibitemOpen
  \bibfield  {author} {\bibinfo {author} {\bibfnamefont {J.~M.}\ \bibnamefont
  {Worlock}}\ and\ \bibinfo {author} {\bibfnamefont {P.~A.}\ \bibnamefont
  {Fleury}},\ }\href@noop {} {\bibfield  {journal} {\bibinfo  {journal} {Phys.
  Rev. Lett.}\ }\textbf {\bibinfo {volume} {19}},\ \bibinfo {pages} {1176}
  (\bibinfo {year} {1967})}\BibitemShut {NoStop}%
\bibitem [{\citenamefont {Ashida}\ \emph {et~al.}(2020)\citenamefont {Ashida},
  \citenamefont {Imamoglu}, \citenamefont {Faist}, \citenamefont {Jaksch},
  \citenamefont {Cavalleri},\ and\ \citenamefont {Demler}}]{Ashida2020}%
  \BibitemOpen
  \bibfield  {author} {\bibinfo {author} {\bibfnamefont {Y.}~\bibnamefont
  {Ashida}}, \bibinfo {author} {\bibfnamefont {A.}~\bibnamefont {Imamoglu}},
  \bibinfo {author} {\bibfnamefont {J.}~\bibnamefont {Faist}}, \bibinfo
  {author} {\bibfnamefont {D.}~\bibnamefont {Jaksch}}, \bibinfo {author}
  {\bibfnamefont {A.}~\bibnamefont {Cavalleri}},\ and\ \bibinfo {author}
  {\bibfnamefont {E.}~\bibnamefont {Demler}},\ }\href@noop {} {\bibfield
  {journal} {\bibinfo  {journal} {Phys. Rev. X}\ }\textbf {\bibinfo {volume}
  {10}},\ \bibinfo {pages} {041027} (\bibinfo {year} {2020})}\BibitemShut
  {NoStop}%
\bibitem [{\citenamefont {Hübener}\ \emph {et~al.}(2020)\citenamefont
  {Hübener}, \citenamefont {Giovannini}, \citenamefont {Schäfer},
  \citenamefont {berger}, \citenamefont {Ruggenthaler}, \citenamefont {Faist},\
  and\ \citenamefont {Rubio}}]{Hannes2020}%
  \BibitemOpen
  \bibfield  {author} {\bibinfo {author} {\bibfnamefont {H.}~\bibnamefont
  {Hübener}}, \bibinfo {author} {\bibfnamefont {U.~D.}\ \bibnamefont
  {Giovannini}}, \bibinfo {author} {\bibfnamefont {C.}~\bibnamefont
  {Schäfer}}, \bibinfo {author} {\bibfnamefont {J.}~\bibnamefont {berger}},
  \bibinfo {author} {\bibfnamefont {M.}~\bibnamefont {Ruggenthaler}}, \bibinfo
  {author} {\bibfnamefont {J.}~\bibnamefont {Faist}},\ and\ \bibinfo {author}
  {\bibfnamefont {A.}~\bibnamefont {Rubio}},\ }\bibfield  {journal} {\bibinfo
  {journal} {Nat. Mater.}\ }\href {https://doi.org/10.1038/s41563-020-00801-7}
  {10.1038/s41563-020-00801-7} (\bibinfo {year} {2020})\BibitemShut {NoStop}%
\end{thebibliography}
\end{document}